Focus Section, The Chimeric Mind

When Strings Tug at Algorithm: Human-AI Sovereignty and Entanglement in Nomadic Improvisational Music Performance as a Decolonial Exploration

Joshua Nijiati Alimujiang

Joshua Nijiati Alimujiang (researcher), Sony Computer Science Laboratory / École Polytechnique (Paris), 6 Rue Amyot, 75005 Paris, France. Email: alimujiang.nijiati@polytechnique.edu. ORCID: 0009-0004-3887-4288.

**Abstract**

As emergent artificial intelligence technologies increasingly assert roles as assistants within intangible cultural heritage contexts, researchers and artists observe existing questions on the theme of agency negotiation, cultural resistance, and technical critique. This research interrogates power dynamics in human-AI sovereignty and entanglement for nomadic improvisational *Dutar* performance, a living cultural heritage through a long-necked lute from the Central Asia region. To investigate tensions between human agency and computational hegemony, the researcher and artists examined and iterated a feedback workflow that captures live performance data, processes digital transformations, and creates a real-time interactive art experience via immersive environments. Empirical data from artists and audience reveal modulations where musicians selectively embrace or reject algorithmic suggestions to preserve creative identity. The author concludes that decolonial potential requires redesigning tools or systems for cultural survivance, where technology becomes not merely a feedback environment but a site for decolonial praxis, challenging computational hegemony in digital ecosystems.

## <1> Beyond the Co-creation: Questioning Identity in Human-AI Improvisation

The emerging paradigm of artificial intelligence as a generator sometimes arrives with unexamined colonial baggage, a universalizing claim of mainstream narrative that risks erasing specific cultural ontologies. This framing is particularly fraught within the context of intangible cultural heritage, where the very concept of creation is often deeply relational, non-linear, and bound to tradition. The Dutar, a traditional long-necked lute, holds a place in the musical heritage of Central Asia, celebrated for its unique sound and cultural resonance. As digital technologies increasingly permeate the arts, there is still a burgeoning gap in exploring how these tools are entangling with traditional artistic practices to discover possibilities of new expression and critique. Provoked by retrieving and discoursing papers with Dutar artists, this work began as a decolonial exploration of questions on human-AI sovereignty and entanglement.

The debate surrounding AI in art often celebrates its technical capabilities, real-time data processing and generative output, as inherently transformative for artistic practice [1]. This narrative is exemplified by systems like GROUPTHINK, which merges physiological feedback with AI to enable remote audience co-creation, proposing a future of distributed agency within an internet of neurons" [2]. Similarly, projects like LuminAI investigate embodied co-creativity, framing the human-AI relationship as one of negotiation within physical improvisation [3]. While such research valuably expands the expressive toolkit [4], it frequently operates within a techno-utopian framework that presumes collaboration is an end in itself. Prevailing co-creation frameworks may risk obscuring the colonial logics embedded within AI systems and miss questioning the underlying power dynamics and cultural assumptions embedded in the AI itself. When uncritically accepted as impartial mediators, AI systems operationalize structural biases via homogenized datasets, rendering invisible both cultural erasures and ideological impositions inherent in co-creative processes.

During the retrieval and investigation, collaborative performance artists considered that the recent integration of immersive technologies has prompted a critical reevaluation of artistic boundaries and authenticity paradigms. In this study, immersive technology is referred to and used as an umbrella term covering virtual reality (VR), augmented reality (AR), mixed reality (MR), and extended reality (XR) technologies [5]. 'Nature: Metaphysics + Metaphor' introduces an art experience that shows innovation in combining painting, sculpture, land art, and performance art with digital technology to investigate MR persistence, feedback, and visualization by exploring the relationship between human and nature [6]. 'NatureBlendVR' utilizes XR technology and bio-responsive elements to enhance emotional regulation and cognitive performance by integrating physical and virtual components in a nature-inspired environment [7]. Crucially, Wang and Martin bridge this discourse toward performative contexts with 'AI See, You See', introducing novel AR visualization techniques that render AI's compositional logic tangible during real-time musical collaboration [8]. Their auto-ethnographic findings reveal how such transparency scaffolds shared agency yet simultaneously reveals tensions around AI's creative presence, an essential precursor to understanding sovereignty negotiations explored herein. Complementary explorations include Mochizuki et al.'s neurodiverse avatar-mediated Noh performances demonstrating XR's capacity for embodied renegotiation of creative roles [9], alongside motion-capture historiographies transforming Angel Island narratives by migratory gesture-projection dialogues [10]. However, some renegotiations are seldom examined through a decolonial lens, especially how these technologies entangle with and potentially disrupt culturally specific, non-Western forms of knowledge and practice.

Revealed via first-hand and second-hand investigations with Dutar artists, this research is going to deliberately enter the space for asking how to review the entanglement itself as a site of critical scrutiny. After a series of interviews and technological workshops, the focus intervention is on mixed reality and immersive screens. This specific choice creates a spatially anchored and visible arena for the human-AI encounter. The MR environment allows the Dutar performer to engage with AI-generated visual narratives in real time, while an iterated design constraint, ensuring the view of hand positions remains unobstructed, physically safeguards the sovereignty of the performer's embodied technique. The audience, in turn, perceives this dialectic not as a pure fusion but as a mediated intersubjectivity on the immersive screen.

## <2> Deconstructing the Hegemonic: Situated and Localized Praxis as Resistance

This artistic research aims to examine how improvisational music performance becomes a contested site for observing and mediating power between human creativity and computational systems. Established on interviews, reviews, and critical analysis with artists, research questions probe the complicated interdependencies where artistic intention intersects with algorithmic influence: (RQ1) How do Dutar performers negotiate artistic sovereignty when their improvisations trigger real-time AI-generated visual narratives, and how does this impact their creative identity? (RQ2) When the algorithmic gaze dictates the visual aesthetics of improvisation, can cultural specificity ever transcend the gravitational pull of algorithmic universalism, and what creative acts might rupture this hegemony? (RQ3) Where do dominant paradigms of 'optimized' technical integration and interactive design inherently favor machine efficiency over human expression, and what slow media approaches could reclaim improvisational temporality?

Collectively, these questions form interlocking interventions: RQ1's micro-level agency struggles necessitate RQ2's macro-cultural resistance, while RQ3's subversion of techno-temporal constraints enables the emancipatory entanglement sought by both. To materialize this critical framework, research methodology enacts entangled agencies through five integrated phases, as Fig. 1.

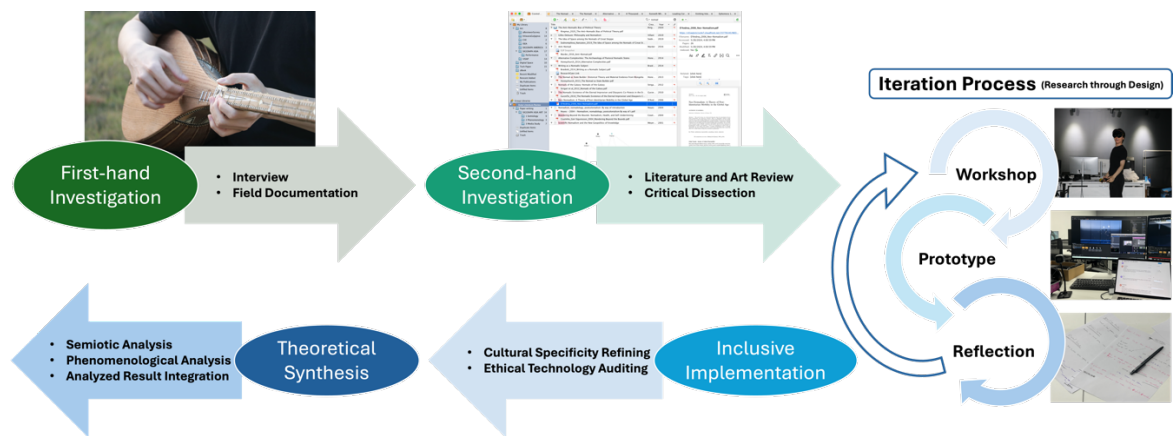

Fig. 1. Joshua Nijiati, Diagram of Artistic Research, digital graphics, dimensions variable, 2025. Artistic research process in this study includes five main procedures from investigation to experiment, and then analysis. In the iteration process, artists practiced diverse immersive technologies, then prototyped and reflected with researchers over 8 cycles. (© Joshua Nijiati.)

*Alt Text:* A flowchart diagram illustrating the five-phase, cyclical research methodology: First-hand Investigation (interviews, field documentation) leads to Second-hand Investigation (literature review, critical dissection), which feeds into an Iteration Process (research through design). Then, it informs Inclusive Implementation (cultural specificity refining, ethical auditing) and culminates in Theoretical Synthesis (semiotic and phenomenological analysis).

Foundationally, first-hand investigation established cultural baselines for Dutar improvisation, revealing communal and temporal agency as counterpoints to tech mediation, directly grounding RQ1's sovereignty inquiry in embodied practice. Critically, second-hand investigation exposed colonial biases in AI tools, necessitating resistance in design, providing the critical infrastructure for RQ2's confrontation with algorithmic hegemony. Reflexively, exploratory iteration mapped agency boundaries through participatory prototyping, where RQ3's slow-media principles materialized via iteration and retrospective analysis that exposed cultural risks in algorithmic interpretation. Operationally, inclusive implementation embedded survivance via curated datasets and ethical auditing, localizing computational aesthetics, operationalizing RQ2's cultural resistance within technical systems. Synthetically, theoretical synthesis converged semiotic-phenomenological analysis with decolonial theory, reframing immersive environments as contested territory for the agential reconfiguration central to all three RQs.

First-hand investigations began with semi-structured interviews of Dutar artists, mapping improvisational sovereignty through their accounts of temporal agency, communal storytelling, resistance to technical mediation, etc. To establish cultural baselines for developing technology, we conducted focused group discussions probing technology sensitivities and documented them with video and writing. Second-hand investigations then expanded through literature and documents through multiple archive sources, including but not limited to Leonardo, AI and Society, Technoetic Arts, Digital Creativity, SIGGRAPH, and VISAP. Crucially, iterative exploration transformed workshops into sites of possibility exploration: after each session, researchers and artists jointly reviewed video documentation and emergent tensions, collaboratively defining problems and prototyping solutions. Over eight cycles, this mediation cultivated balanced expression: artists selectively ceded visual expression control to computational systems while retaining cultural authorship through providing original photography as part of datasets and fine-tuning the generative process by adding or blocking prompt keywords. The final prototype, selected as our case study, crystallizes this hard-won symbiosis and sovereignty between human intuition and machine logic.

### <3> Mediating the Chimeric: Artistic and Cultural Sovereignty as Resilience

Synthesizing insights from phased investigation, the first-hand investigation revealed that Dutar improvisation thrives on temporal elasticity and communal dialogue, elements threatened by digital mediation, which feature in drawing attention and distracting concentration. Concurrently, the second-hand investigation exposed latent colonial biases in mainstream AI datasets, which risked homogenizing culturally specific sonic semantics. These critical tensions were then prototyped, practiced, and negotiated during the iterative process, where artists collaboratively mapped boundaries for human-machine agency through experiments and exercises. This tripartite process crystallized into a core vision: to discover and examine an interactive system where AI and immersive technologies augment, rather than overwrite, artistic and cultural sovereignty.

To implement this process into critical and open-ended explorations and experiments, specifically addressing how performers negotiate native consciousness, resist algorithmic universalism, and reclaim improvisational temporality, the researcher and artists iteratively developed the interactive system integrating AI and immersive technologies for traditional dutar performance based on the needs that artists proposed. Building upon studies that examined the role of AI in music composition and performance [11-12], the research team reconceptualizes visualization as a site for sovereignty negotiations during live performance. By designing a workflow capturing live Dutar performance sound, the researcher visualizes soundwave using TouchDesigner, producing dynamic imagery that serves as input for the StreamDiffusion model, which generates narrative content for video frames based on AI interpretations of the music and pre-trained cultural documentation in visual and literal forms. These AI-generated contents are then integrated into the MR environment using Unity, where they are experienced through headsets by performers as algorithmic companionship and displayed on the immersive screen for audiences as mediated intersubjectivity, creating a real-time feedback loop that influences the musician's improvisation. This recursive entanglement loop generates emergent narrative strata through machine interpretation, simultaneously constructing negotiated agencies between performer-AI entanglement and distributed spectatorship across immersive displays, revealing how AI reconstitutes artistic risk-taking within culturally grounded improvisation. The implementation workflow depicted in Fig. 2 operationalizes the technological framework, detailing the symbiotic loop from capturing live sonic gestures to mediating AI-generated visual narratives within the mixed reality environment.

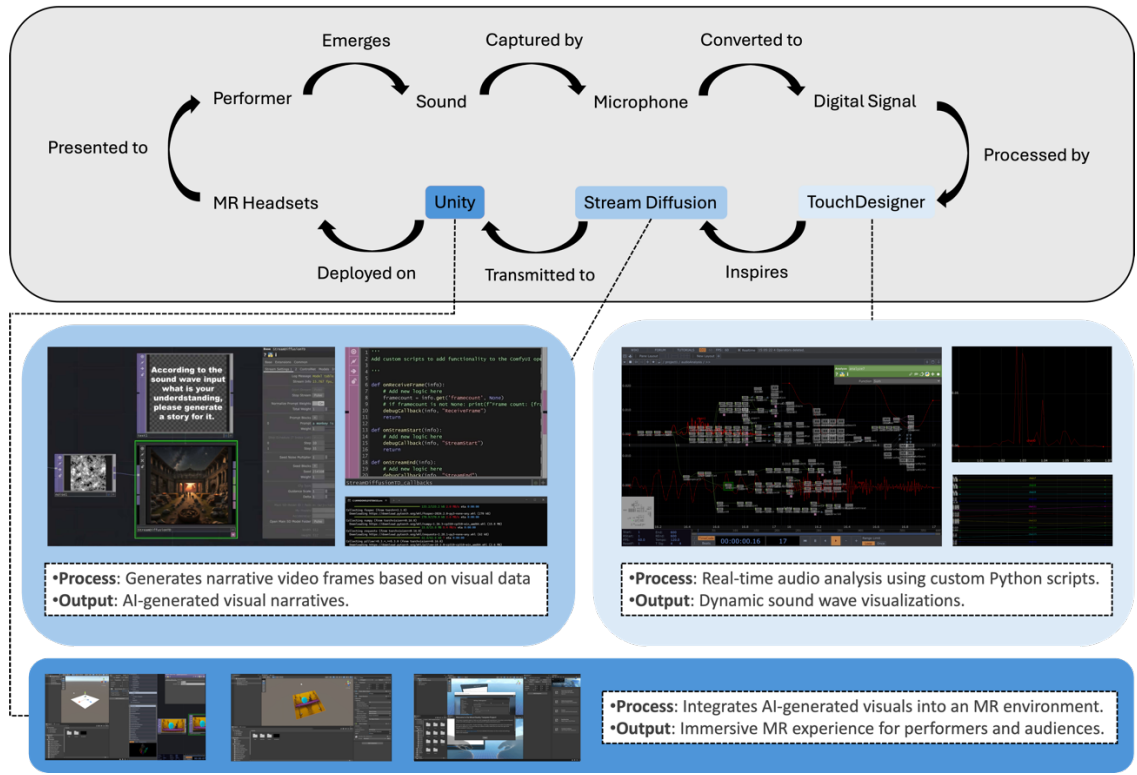

Fig. 2. Joshua Nijiati, Diagram of Technological Implementation, digital graphics, dimensions variable, 2025. Implementation diagram as a cyclic loop for real-time interaction performance, excited by the sound of the performer playing the Dutar. (© Joshua Nijiati.)

*Alt Text:* A diagram illustrates the real-time workflow of an interactive mixed reality system for musical performance. The process begins with the Performer creating sound, which is captured by the microphone. The analog sound is converted to a digital signal and processed by TouchDesigner, which is then provided as the generating reference base for the Stream Diffusion model. The outputs are then transmitted to Unity to integrate them into a cohesive Mixed Reality environment. This MR experience is finally deployed on headsets and presented to both the performer and the audience, creating a real-time feedback loop.

In the sound capture and processing, Dutar performances were collected using a Neumann TLM 103 condenser microphone. The microphone was connected to a Focusrite Scarlett 2i2 audio interface to ensure high-quality digital sound recording. The recorded audio files were imported into TouchDesigner, a node-based visual programming environment. Custom Python scripts were developed to analyze audio features such as frequency, amplitude, and tempo. These scripts generated real-time visualizations that varied in color, shape, and motion, effectively translating the music's dynamics into visual form, which is selectable in the following steps by artists.

With the purpose of investigating AI interpretation and narrative generation in decolonial exploration, the Stream Diffusion model, based on a diffusion process for generating content, was adapted for this project. The model had been pre-trained on datasets including visual and literary motifs. Fine-tuning involved training the model on additional datasets featuring Central Asian cultural imagery and literature to align Dutar's cultural context. Documentation and content sources are various in type, as shown in Fig. 3. Visual patterns from TouchDesigner were exported as high-resolution image sequences. These sequences were input into the Stream Diffusion model, which processed them to generate frames of a narrative content video as the lower half of Figure 3. Artists have the option to regenerate during the performance.

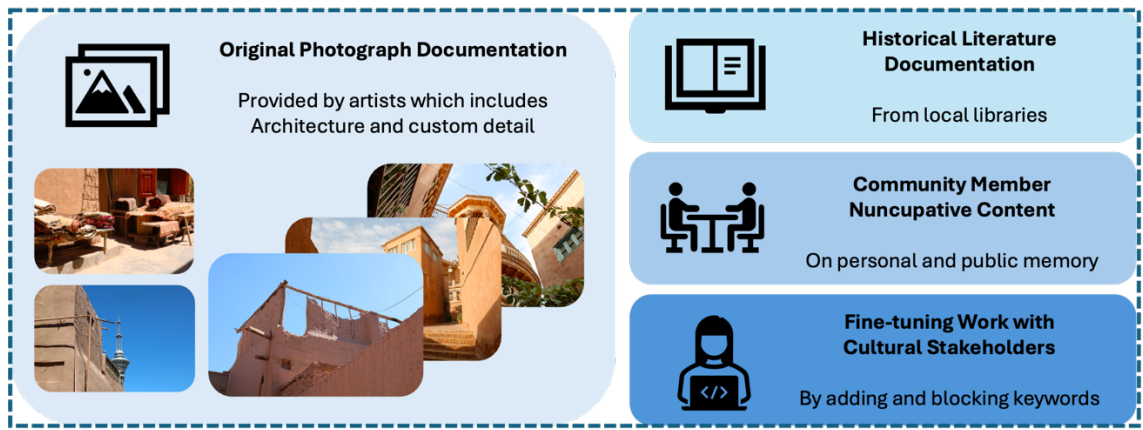

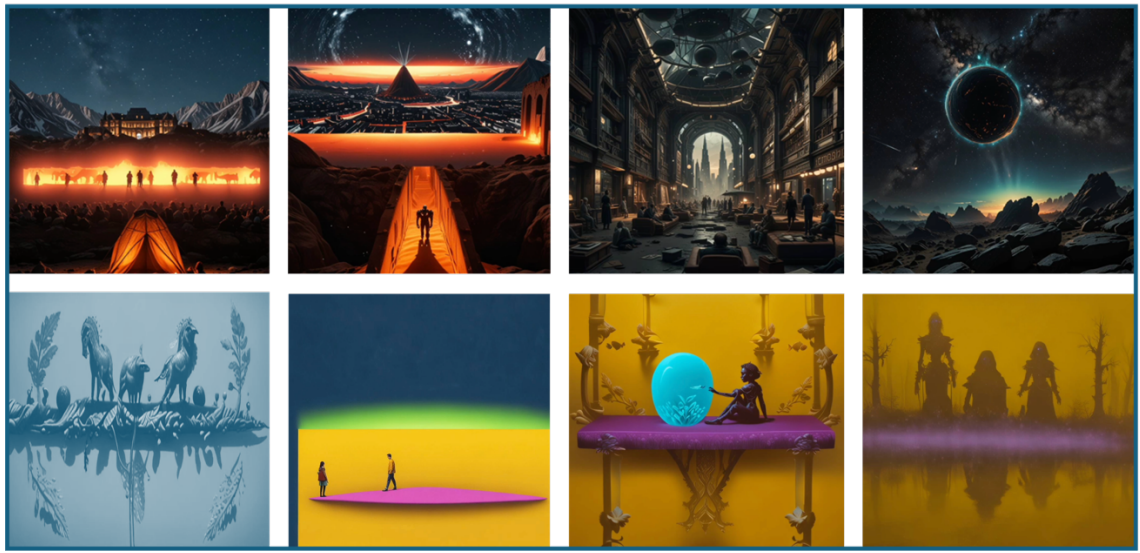

Fig. 3. Joshua Nijiati, Diagram of Providing Cultural Database Sources, digital graphics, dimensions variable, 2025. From the native photograph, historical literature, nuncupative record, and fine-tuning process with cultural stakeholders, the result aims for localized and decolonial experiments for probing the boundaries of sovereignty and entanglement. (© Joshua Nijiati.)

*Alt Text*: The diagram outlines the composition of a culturally informed dataset, created by synthesizing original photographs, historical texts, and community oral histories. This curated collection was then used to fine-tune an AI model, employing targeted prompt engineering to align its generative output with specific cultural contexts and narratives.

For the integration into the MR environment, Unity 2021.2 was used to develop the MR application, leveraging its XR Interaction Toolkit for AR functionality. The AI-generated video frames were imported into Unity and layered with adjustive elements to create a cohesive MR experience. Visual effects and shaders were applied to blend the digital content with the real world. Meta Quest Pro and HoloLens headsets were both deployed for MR display. The performer interface allows users to manipulate the visuals. The experience was designed to be intuitive, requiring minimal learning for performers to adapt and watch the AI-generated content in an immersive environment, in which performers can also observe the audience and reality. It is the workshops' result, leading to an interface that does not cover or influence the majority of the real world, as improvisational performers' needs in reacting to audience real-time feedback.

The real-time feedback loop was implemented using audio processing and API scripting. The system continuously captured live audio input, which was analyzed and used to update the visual content dynamically. The loop ensured that changes in the performance, such as tempo variations or dynamic shifts, were reflected in the visuals. Performers were equipped with Meta Quest Pro or HoloLens headsets by their preference, allowing them to see the real-time visualizations. This interaction encouraged improvisation, as artists responded to visual cues. Audience members viewed the performance with screens displaying the AI-generated narrative content, as Fig. 4, enhancing their engagement by showing the symbiotic relationship between sound and visuals during the performance.

Initial tests and evaluations were conducted with Dutar artists in the form of workshops to collect their advice, followed by audience groups to watch the performance. Feedback was collected through surveys and interviews, focusing on the system's usability, the coherence of the visual-music interaction, and overall satisfaction. Metrics such as frame rate stability, audio-visual synchronization, and performer interaction were monitored to ensure further analysis and improvement. Adjustments were made based on test results to enhance performance. After times of iterative tuning, this version of the system achieved both artist and researcher expectations.

### <4> Hearing the Entanglement: Performer and Audience Analysis as Dialectic

As McCormack et al. note, the collaboration between human artists and AI systems often leads to new creative horizons, expanding the possibilities for artistic expression and audience engagement [13]. This project builds on the exploration of discovering the intersection of technology and tradition, inspired by the early experiments of Nam June Paik in video art and the generative AI program AARON systems by Harold Cohen [14]. Epistemologically, this work is dedicated to engaging with debates about the role of machines in creative processes, echoing

Walter Benjamin's concerns about the loss of "aura" in mechanically reproduced art while also embracing the potential for new forms of authenticity in digital [15].

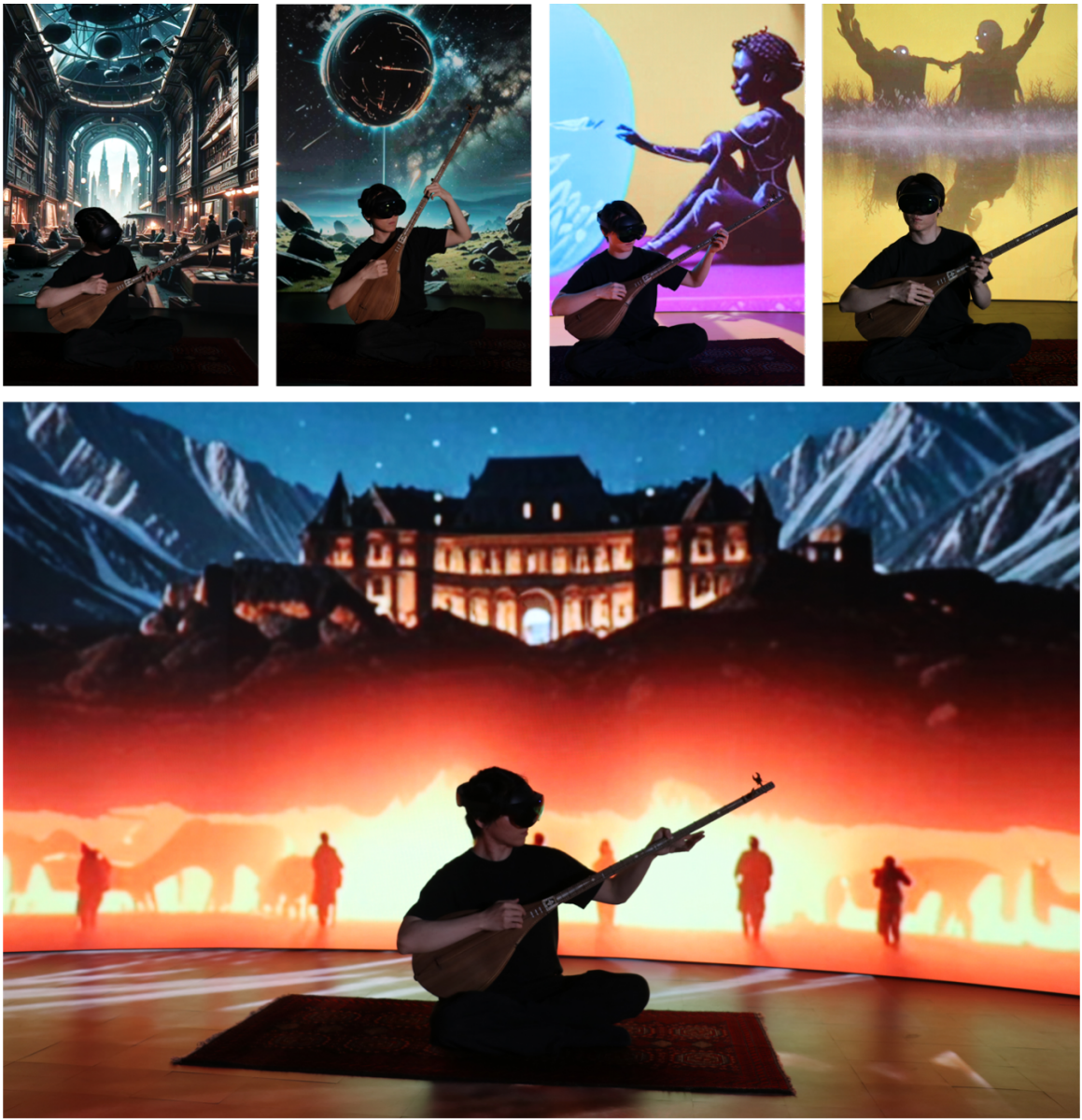

Fig. 4. Joshua Nijiati, Photograph of Performance Site, digital graphics, dimensions variable, 2025. As the Dutar performer plays, the real-time narrative content is displayed in the headset and immersive screen. (© Joshua Nijiati.)

*Alt Text*: Five photos document a live musical performance from the audience's point of view, where a traditional musician interacts with an immersive visual landscape. The collection illustrates the dialogue between the physical musician and the digital world they co-create.

Performer X described the lasting influence on their improvisational performing due to the dynamic visualizations accompanying their music. "The visuals created by the AI expanded my imaginative horizon," remarked the performer. "It was like having a visual partner on stage, inspiring my improvisation." The performer found that the visual feedback enhanced the musical expression and introduced new dimensions to performance styles. Initially, performer Y faced challenges adapting to the new device; therefore, the researcher provided various equipment and technologies for the performer to experience and provide feedback. "Once I got the hang of this MR headset, it felt like a seamless extension of my instrument!" This result underscores the performers' artistic exploration and practice for the fusion of immersive technology and tradition, aligning with studies on human-AI collaboration in creative processes [16]. It, to some extent, mirrors Marshall McLuhan's concept of media as extensions of the human body, where technology becomes an integral part of the artist's creative toolkit [17].

The audience's reflections further discuss the project's potential for engagement and experience. About fifty-eight percent of attendees noted that the AI-generated narrative content deepened the synesthesia connection to the music. "Seeing the music come to life visually was a mesmerizing experience, and it deepened my connection to the performance in a way I hadn't expected," commented audience member A. This immersive experience allowed the audience to perceive the music in synesthetic ways, bridging the gap between auditory and visual senses, a phenomenon explored in research on multisensory art experiences [18]. Another audience member B from the local community, mentioned, "I was initially skeptical about mixing traditional music with digital art, but the result was a beautiful fusion that respected the essence of the dutar while bringing something new to the table." Such reflections indicate the acceptance and appreciation of blending cultural heritage with modern technology, as discussed in studies on the revitalization of traditional arts through digital innovation [19]. It aligns with Lev Manovich's argument that new media art is not to replace traditional forms but rather recontextualize them, creating hybrid expressions that resonate with contemporary audiences [20].

Building on the last participant feedback, we redeveloped an implementation workflow (Fig. 5a) that enables audience members to influence improvisational performances by selecting environmental sounds through interaction with symbolic objects placed in front of the performer. After the audience selects a representative 3D-printed object which is embedded sensors, the ambient sound will be triggered, as the sound of rain in the forest, bonfires in the desert, leaves in the wind, etc. This modified approach aligns with theories of participatory art articulated by Bishop, who emphasizes the transformative potential of audience agency in redefining artistic experiences [22]. Furthermore, we redesigned the Unity-based MR environment to incorporate dynamic adaptability features, allowing performers to adjust not only the scale of AI-generated visual content but also the spatial orientation and transparency of display windows. These modifications enhance system responsiveness to diverse environmental lighting conditions, addressing key challenges in MR visibility identified in studies by Grubert et al. [23]. The adjustable parameters can enable real-time calibration between projected digital elements and physical staging requirements, creating a more seamless integration of mixed reality components as described in Milgram's reality-virtuality continuum [24].

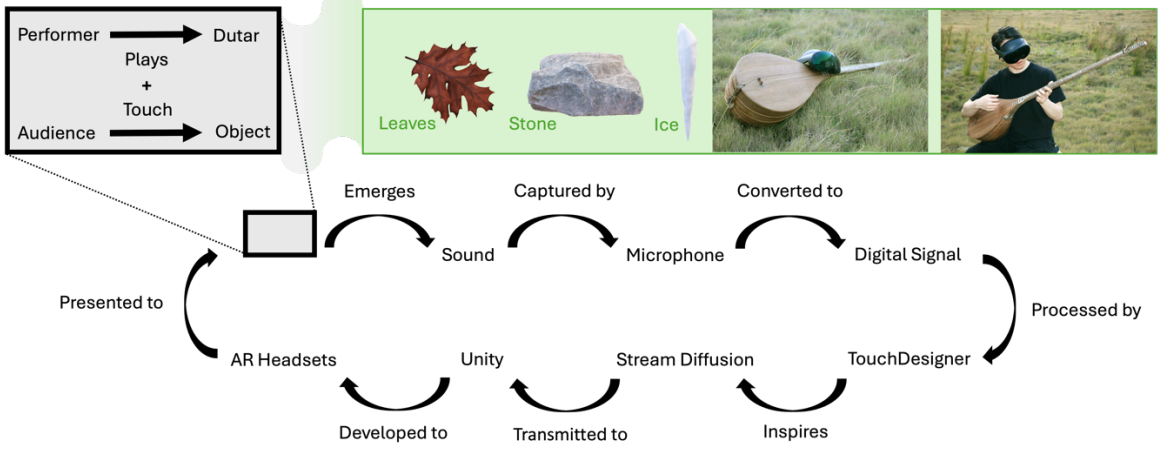

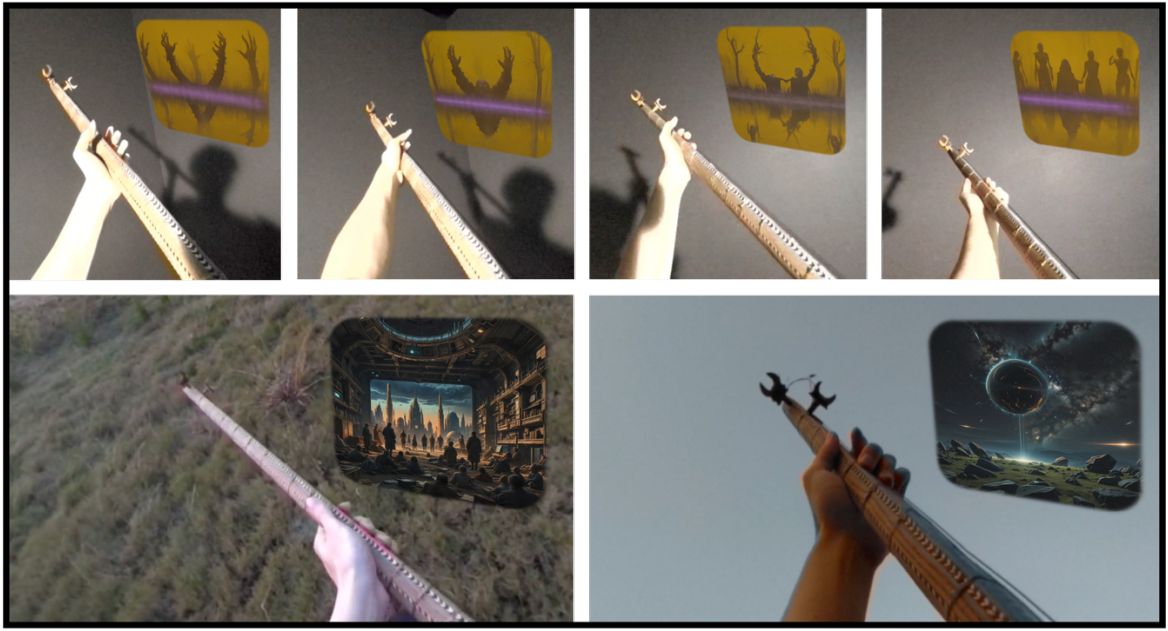

Fig. 5. Joshua Nijiati, Diagram of the Iterated Version of Technological Implementation, digital graphics, dimensions variable, 2025. Implementation workflow and refined performer user interface after reconsidering the public performance's feedback. (© Joshua Nijiati.)

*Alt Text*: A diagram illustrating both the participatory implementation pattern and the adjustable user interface. Above is the improved workflow with symbolic object samples. Arrows indicate that the Audience can touch these objects to trigger corresponding environmental sounds like wind, earth, or water. Below are performance perspectives of adjustable performance interfaces.

Although the futuristic visual direction was guided by artist Z for one music set that discussed the mythology about the universe based on fable allusions in the nomadic community. But for this contrast that the artist explored, there was also a call for more cultural context details in the visuals that ensure they resonate more closely with the instrument. "While the visuals that combined tradition and futurism were beautiful, I felt they could have drawn more directly from the cultural heritage of the dutar," suggested a cultural expert C in the audience. This feedback points to the potential for integrating more culturally specific elements into the visual design and model training, enhancing authenticity and cultural resonance, as emphasized in studies on the importance of cultural context in digital art [25]. This critique aligns with Edward Said's postcolonial perspective, which emphasizes the need to balance cultural narratives in the face of technological homogenization [26]. However, it can still be seen as an experimental attempt at discovering the boundary of audience acceptance on this subject.

## <5> Observing the Uncertainty: Nomadic and Quantum Consciousness as Critique

Drawing upon semiotic analysis rooted in Peircean triadic models [27], where musical signals function as indexical signs pointing to cultural epistemologies, current findings in this research reveal that Dutar performers assert sovereignty by embedding traditional techniques within training visual and literal datasets as deliberate symbolic anchors against algorithmic homogenization. The semiotic framing enabled tangible agency retention during real-time visual generation, evidenced when artists overrode AI suggestions, misinterpreting Nomadic poetic meters, a phenomenological manifestation of creative identity preservation where corporeal mastery mediated computational logic. Confronting RQ2's tension between cultural specificity and algorithmic universalism, we observed rupture tactics emerging from semantic dissonance: mainstream-tuned visual results included dissonance with subjective orientation that exposed latent colonial biases in generative models, whereas diverse datasets that co-trained with culturally related stakeholders display detailed nuances in native perspectives, amplifying sometimes performer-audience resonance to community memory and identity in variety. These interventions materialized Edward Said's critique of Orientalist techno-cultural erasure through embodied resistance, wherein musicians reclaimed narrative ownership via pretrained visual sets and blocked bias keywords before StreamDiffusion inference cycles. Addressing RQ3's critique of efficiency paradigms, phenomenological unpacking of temporality could reveal how optimized systems compress improvisational duration into discrete computational steps. Such fragmentation disrupts what Sudnow terms the ways of the hand, the lived temporality of embodied expertise cultivated through years of practice, reducing organic decision-making cycles into mechanical transactions [28]. Synthesizing these strands, we posit immersive interfaces not merely as technical scaffolds but as contested semiospheres where slow media principles facilitate what Margaret Boden terms subordinate collaboration, AI generating structural possibilities while human intention governs semantic validity, thus actualizing decolonial praxis at the human-machine nexus.

The notion of nomadic consciousness emerges as a pivotal framework for understanding the fluidity and adaptability of identity within artistic practices in this research, particularly in the context of Dutar performances intertwined with artificial intelligence. This concept resonates deeply with the principles of quantum theory, which emphasizes the inherent unpredictability and interconnectedness of various states of existence. As subatomic particles inhabit a realm of probabilities until observed, the identities of performers are not fixed; rather, they dynamically evolve in response to their interactions with AI technologies and audience feedback. This phenomenon can be described as a quantum consciousness, wherein multiple realities coexist and artists navigate their personal identity and cultural legacies amidst the complexities introduced by algorithmic systems.

By viewing creativity through the lens of quantum consciousness, artists can deconstruct the limitations imposed by conventional paradigms that seek to standardize diverse expressions under the guise of efficiency. This critique aligns with Wendt's assertion in 2005 that a quantum perspective allows for a richer understanding of social phenomena, facilitating deeper engagement with the intricate interplay of agency and interaction [29]. Wendt further elaborates in 2015 on the implications of a quantum mind for social science, highlighting the potential for approaches that embrace uncertainty and fluidity. This expanded understanding of nomadic and quantum consciousness can reframe the conversation around artistic sovereignty and cultural identity by emphasizing the significance of embracing ambiguity in creative processes.

In summary, this research has illuminated complex interplays between agency, cultural identity, and technological mediation, revealing how Nomadic artists assert their sovereignty and explore the entanglement amidst the encroachment of algorithmic influences. By embedding traditional heritage into the training datasets for AI systems, the researcher and musicians have created a framework that critiques the homogenizing tendencies of digital technologies. The feedback gathered from performers and audience members highlights the importance of iterative processes that value cultural specificity, suggesting that situated integration of technology in artistic practices requires individualized approaches explored with artists. Moreover, this study envisions and endeavors to discover possibilities for audience engagement as well, where participants can influence performances through interactive experiences. By reimagining the role and boundary of technology, it found its way for an inclusive and participatory model of native art creation. As navigating in the complex landscape of digitalization, it becomes increasingly vital to ensure that artistic sovereignty is upheld, allowing cultural narratives to thrive sustainably in the face of rapid technological advancement. This research, in the end, advocates for slow media approaches with iterative methodologies that prioritize human expression and creativity over mere technology-centered theory by in-person contact with the serving group of people to honor and facilitate the voices of the audience and artists from various communities.

**Biographical Information**

Joshua Nijiati Alimujiang is a researcher from Sony Computer Science Laboratory (Paris). He's also doing studies in computer vision and artificial intelligence at École Polytechnique, Paris.